\documentclass[journal]{IEEEtran}
\ifCLASSINFOpdf
\else
\fi
\usepackage{tabularx}
\usepackage{makecell}
\usepackage{graphicx}
\usepackage{subfig}
\usepackage{caption}
\usepackage[font=footnotesize]{caption}
\usepackage{amsmath}
\usepackage[switch,pagewise]{lineno} 
\usepackage{lipsum} 

\usepackage{xcolor}
\usepackage{soul}

\begin{document}
%
\title{An Investigation of Drivers' Dynamic Situational Trust in Conditionally Automated Driving}
%
%
%

\author{{Jackie Ayoub, Lilit Avetisyan, Mustapha Makki, Feng Zhou}

\thanks{J. Ayoub, L. Avetisyan, M. Makki and F. Zhou are with the Department of Industrial and Manufacturing, Systems Engineering, The University of Michigan-Dearborn, Dearborn, 4901 Evergreen Rd. Dearborn, MI 48128 USA (e-mail: \{jyayoub, lilita, mumakki, fezhou\}@umich.edu).}
\thanks{Manuscript received March 11, 2021; revised xxx xx, 2021.}}

%
%

\markboth{IEEE Transactions on Human-Machine Systems,~Vol.~xx, No.~xx, Month~2021}%
{Ayoub \MakeLowercase{\textit{et al.}}: An Investigation of Drivers' Dynamic Situational Trust in Conditionally Automated Driving}
%

\maketitle

\begin{abstract}
Understanding how trust is built over time is essential, as trust plays an important role in the acceptance and adoption of automated vehicles (AVs). This study aimed to investigate the effects of system performance and participants' trust preconditions on dynamic situational trust during takeover transitions. We evaluated the dynamic situational trust of 42 participants using both self-reported and behavioral measures while watching 30 videos with takeover scenarios. The study was a 3 by 2 mixed-subjects design, where the within-subjects variable was the system performance (i.e., accuracy levels of 95\%, 80\%, and 70\%) and the between-subjects variable was the preconditions of the participants' trust (i.e., overtrust and undertrust). Our results showed that participants quickly adjusted their self-reported situational trust (SST) levels which were consistent with different accuracy levels of system performance in both trust preconditions. However, participants' behavioral situational trust (BST) was affected by their trust preconditions across different accuracy levels. For instance, the overtrust precondition significantly increased the agreement fraction compared to the undertrust precondition. The undertrust precondition significantly decreased the switch fraction compared to the overtrust precondition. These results have important implications for designing an in-vehicle trust calibration system for conditional AVs.

\end{abstract}


%
\IEEEpeerreviewmaketitle

\section{Introduction}
%
%
%
%







\IEEEPARstart{T}echnological advances in the automotive industry are bringing automated vehicles (AVs) closer to road use. AVs offer numerous benefits and, most importantly, the ability to decrease road crashes by almost 90\% \cite{xu2017}. However, undesirable trust levels in AVs (i.e., overtrust and undertrust) can potentially diminish these benefits. For instance, one of the leading causes of recent AV crashes (e.g., Tesla's fatal crash in Florida \cite{rice2019} and the Uber AV crash in Arizona \cite{kohli2019}) was drivers' overtrust in their AVs' capabilities. Takeover transitions should be promptly and safely handled when AVs reach their functional limit \cite{zhang2019determinants,cao2021towards}. Although these crashes probably occured due to overtrust in the capabilities of conditional AVs (i.e., Society of Automotive Engineers (SAE) Levels 2-3), the crashes might reflect a negative first impression with respect to public opinion about AV safety and capabilities. Even for fully AVs, a survey conducted by AAA showed that three out of four Americans did not trust AVs \cite{edmonds2019}. Wortham et al. \cite{wortham2017} showed that the lack of public trust is a critical factor in the adoption of AVs, especially with the increasing reports of AV failures. 

Therefore, an appropriate level of trust between humans and AVs is needed to improve safety and acceptance simultaneously. As a result, it is important to understand the definition and different constructs of trust in AVs. Lee and See \cite{lee2004} defined trust as “\emph{the attitude that an agent will help achieve an individual’s goals in a situation characterized by uncertainty and vulnerability}”. Thus, in conditional AVs, takeover transitions influence drivers' trust because of the uncertainty and vulnerability associated with the takeover transition, especially when the driver is performing nondriving-related tasks (NDRTs) during the drive \cite{schwarz2019,hergeth2016,korber2018,hergeth2017}. For example, Hergeth et al. \cite{hergeth2016} investigated the impact of takeover requests (TORs) on participants' trust and found that the first TOR had a strong negative effect on their trust. 

Furthermore, Hoff and Bashir \cite{hoff2015} identified three constructs of trust: dispositional, situational, and learned trust. Dispositional trust reflects the tendency to automation and is mainly affected by age, culture, gender, and personality traits. Situational trust depends on the situation and it can be affected by internal (e.g., emotion, self-confidence, and expertise) and external (e.g., system complexity and performance, task difficulty) variables. Learned trust is influenced by past experience and knowledge about automation. In our previous study \cite{Ayoub2021Modeling}, we aimed to measure the public's dispositional and initial learned trust in AVs through an explainable machine learning model. In addition, we identified important factors affecting dispositional and initial learned trust, such as perceived risks and benefits, and knowledge and emotions about AVs. However, in conditional AVs, it is important to understand dynamic situational trust under the influence of different factors. 

System performance is one of the important factors affecting drivers' dynamic situational trust in AVs \cite{merritt2015well}. In general, a consistently good system performance improves trust and vice versa. However, many studies (e.g., \cite{yin2019}, \cite{okamura2020}) reported the influence of the system performance on participants' overall trust without investigating the trust dynamics. For example, Yin et al. \cite{yin2019} found that participants' situational trust was affected by the model's stated and observed accuracy, but the researchers did not measure the effect of accuracy on the trust dynamics over time.

One reason to investigate the dynamics of situational trust over time is that people can be potentially 'trapped' in two undesirable trust conditions (i.e., overtrust and undertrust), and it takes time to calibrate their trust with varied system performance. For instance, Schwarz et al. \cite{schwarz2019} investigated the effect of varying AV system reliability on participants' trust in conditionally automated driving situations. They found that individuals who did not experience TORs in the initial drive (i.e., in an overtrust condition) did not decrease their trust in AVs in the second drive after experiencing TORs. Hence, it is important to examine the dynamics of situational trust over time. Okamura et al. \cite{okamura2020} showed that, by adaptively presenting simple cues, participants in an overtrust precondition were able to calibrate their trust level dynamically over time.

To understand how situational trust changes dynamically over time, it is important to investigate how both overtrust and undertrust preconditions influence situational trust when trapped in two undesirable trust conditions, trust calibration takes more time. We investigated how situational trust changed dynamically at different system performance levels when participants were under an overtrust or an undertrust precondition. We measured 42 participants' situational trust using self-reported and behavioral measures after they viewed 30 videos of conditional AVs, with and without failures, during a 2 by 3 mixed-subjects experiment. The within-subjects variable was the system performance with three accuracy levels (i.e., 95\%, 80\%, and 70\%), while the between-subjects variable was the trust precondition (i.e., overtrust and undertrust). In summary, the contributions of this study are: 
\begin{itemize}
  \item We simultaneously investigated the effects of system performance (i.e., 95\%, 80\%, and 70\%) and trust preconditions (i.e., overtrust and undertrust) on the dynamic situational trust in conditionally automated driving.
  \item We used both self-reported and behavioral trust measures to understand the dynamic situational trust in conditionally automated driving.
  \item The insights obtained from dynamic situational trust provided important implications for calibrating trust over time in conditionally automated driving. 
\end{itemize}



\section{Related Work}
The concept of trust has increasingly gained attention due to its importance in human-AV interactions. Ayoub et al. \cite{ayoub2019} summarized the factors affecting trust in AVs into three categories: 1) automation-related factors (i.e., uncertainty, reliability, user interface, and workload), 2) environment-related factors (i.e., brand, manufacturer reputation, and weather), and 3) human-related factors (i.e., age, culture, gender, AV Knowledge, personality traits, and experience). Lee and See \cite{lee2004} stated that trust was characterized by uncertainty and vulnerability \cite{lee2004}, and emphasized its dynamic nature. Hoff and Bashir \cite{hoff2015} categorized trust into three categories: dispositional, situational, and learned and identified different factors that formed these trust constructs. They found that  although dispositional trust tended to be stable over time, both situational trust and learned trust could change dynamically. 

Many researchers have investigated the factors associated with different levels of trust. For instance, Ayoub et al. \cite{Ayoub2021Modeling} identified important factors affecting people's dispositional and learned trust in AVs using an explainable machine learning model. They found that an individual's perceived benefits and risks of AVs, excitement about driving, knowledge of AVs, and eagerness to adopt a new technology were ranked the highest. 
Zhang et al. \cite{zhang2019} examined the effects of different factors on initial learned trust in AVs using a technology acceptance model and found that perceived safety risk was negatively associated with initial learned trust, while perceived usefulness was positively associated with initial learned trust. They showed that initial learned trust had a high impact on enhancing AV acceptance. In addition, initial learned trust in AVs increased with more interactive automated driving. For example, Gold et al. \cite{gold2015} found that participants' self-reported (learned) trust was enhanced after experiencing a drive with three TORs. Similarly, Beggiato et al. \cite{beggiato2013} showed that participants' learned trust in adaptive cruise control increased when they were provided with a description of the system capabilities and limitations. However, these studies mostly adopted a snap-shot view that measured trust before and after the experiment without examining the dynamics of trust over time \cite{guo2020modeling}, which is critical to calibrating overtrust or undertrust.

Hoff and Bashir \cite{hoff2015} identified many factors influencing situational trust, including task difficulty, system performance, perceived risks and benefits, driver workload, experience, attentional capacity, and mood. 
However, few researchers investigated the dynamics of situational trust in conditional AVs. Hergeth et al. \cite{hergeth2016} showed that participants' self-reported situational trust (SST) dynamically increased from the first to the eighth TOR. Azevedo-Sa et al. \cite{azevedo2020real} presented a framework for estimating participants' dynamic situational trust due to malfunctions of the AV system. Luo et al. \cite{luo2020trust} examined dynamic situational trust using two variables, i.e., level and source of stochasticity. They found that participants' trust decreased significantly due to AV internal errors (e.g., sensor error) versus external  (e.g., roadblocks) errors. Okamura et al. \cite{okamura2020} examined the effect of the AV system dynamic reliability on participants' situational trust and used trust calibration cues to help improve performance. These studies gained insights into how these variables influenced situational trust. 


To understand how situational trust evolves over time, it is important to measure it properly. 
The majority of existing measures are based on questionnaires, such as the trust scale proposed by Jian et al. \cite{jian2000}, which measures overall trust. One limitation of this scale is that it cannot support the temporal and context-related nature of situational trust. Recently, Holthausen et al. \cite{holthausen2020} suggested a short scale for faster implementation based on the trust model suggested by Hoff and Bashir \cite{hoff2015}. The suggested scale evaluated situational trust using six items, including trust, performance, NDRT, risk, judgment, and reaction. Trust was also reported to be highly related to individual behaviors \cite{lee2004,hoff2015}. For example, Hergeth et al. \cite{hergeth2016} showed that monitoring frequency during NDRTs in conditional AVs was negatively correlated with three constructs of trust. Therefore, both self-reported measures and behavioral measures should be considered when determining how situational trust evolves.

\begin{figure}[bt!]
\centering
\includegraphics[width=1\linewidth]{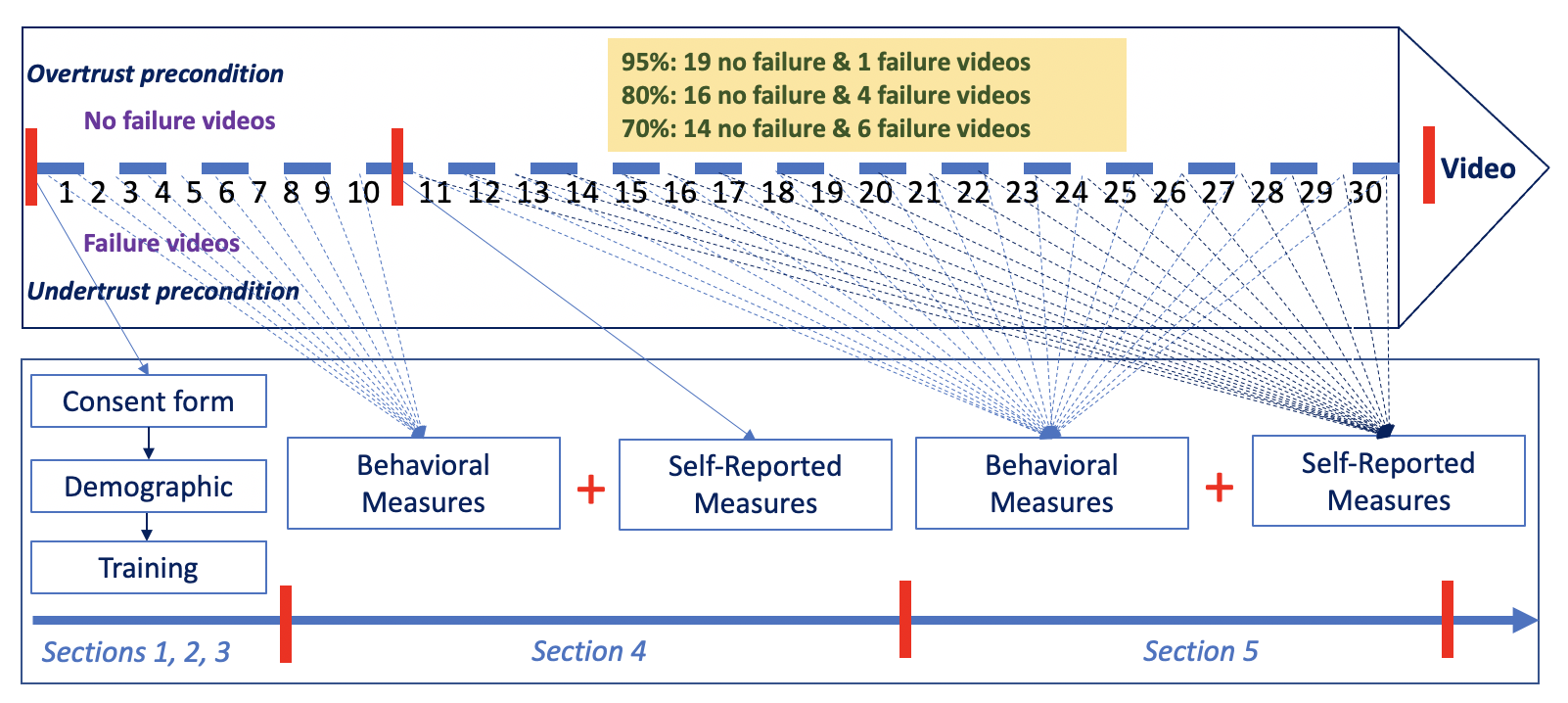}
\caption{Survey procedure.}
\label{fig:procedure}
\end{figure}
\section{Method}

\subsection{Participants}
A total number of 42 participants (22 females and 20 males; M = 25.0 years and SD = 5.4 years) located in the United States participated in this study. All the participants were university students, and each had a valid U.S. driver's license. Participants were randomly assigned to one of the two preconditions (i.e., overtrust and undertrust) of the experiment. This study was conducted in accordance with the ethical requirements of the Institutional Review Board at the University of Michigan (i.e., application number is HUM00190626). Participants were compensated with a \$15 Amazon gift card upon completion of the study and/or partial course credit. The average completion time of one session of the study (i.e., 33 minutes) was similar in the two tested preconditions. 
%


\begin{figure*}[h!]
\centering
\includegraphics[width=0.9\linewidth]{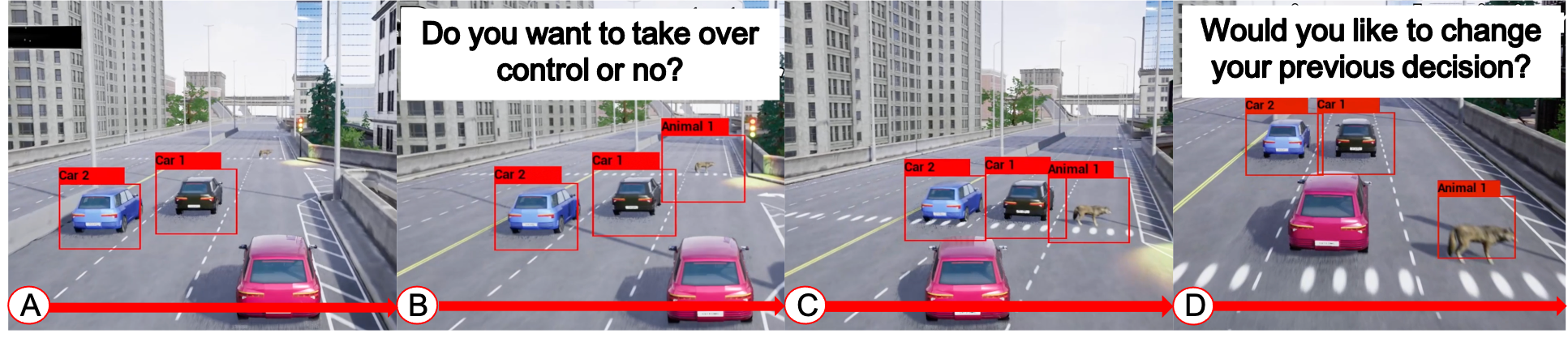}
\caption{Flow of measuring behavioral situational trust during a takeover scenario with no failures.}
\label{fig:videosteps}
\end{figure*}

\begin{table*}[tb]
\centering
\caption{A description of the created scenarios.}
\label{table:scenariodescription}
\begin{tabular}{lccccc}
\hline
\hline
                                    & \textbf{Group 1}                     & \textbf{Group 2}             & \textbf{Group 3}        & \textbf{Group 4}                 & \textbf{Group 5}       \\ \hline
\textbf{Location}                   & Highway or Street                    & Highway or Street            & Highway or crossroad    & Crossroad               & Highway       \\
\multicolumn{1}{c}{\textbf{Target}} & Pedestrian or Animal or construction & Truck or Vehicle             & Vehicle                 & Red light               & Vehicle       \\
\multicolumn{1}{c}{\textbf{Action}} & Road crossing                        & Broken down truck or vehicle & Stopping at a stop sign & Stopping at a red light & Merging lanes \\ \hline
\end{tabular}
\end{table*}

\subsection{Apparatus} 

A survey was designed to evaluate the effects of trust preconditions and system performance on participants' situational trust during takeover scenarios. The survey was developed using Qualtrics (Provo, UT) and administered by two of the authors using Zoom (San Jose, CA). Driving scenarios with TORs were created in a virtual environment using Unreal Engine (Cary, NC). Longitudinal and lateral control was provided, and the maximum speed was set to 50 miles per hour, which mimicked an SAE Level 3 AV. The system was also designed to overtake in some scenarios to successfully avoid obstacles. 

\subsection{Experiment Design}

\textbf{Independent variables.} Our experiment was a 3 (system performance with three accuracy levels: 95\%, 80\%, and 70\%) by 2 (trust preconditions: overtrust and undertrust) mixed-subjects design. The within-subjects variable was the system performance. The between-subjects variable was the trust precondition of the participants, who were shown ten takeover scenarios consecutively with successes (to elicit overtrust) or failures (to elicit undertrust). 

\textbf{Dependent variables.} We measured participants' dynamic situational trust during the experiment using two measures: 1) self-reported measures using the Situational Trust Scale for Automated Driving (STS-AD) \cite{holthausen2020} and 2) behavioral measures to capture how often participants agreed or disagreed with the AV's decision. The STS-AD included six scale items based on the trust model of Hoff and Bashir \cite{hoff2015} (i.e., trust, performance, NDRT, risk, judgment, and reaction). The statements presented in this study to evaluate the SST were as follows: Q1) I trust the automation in this situation, Q2) I would have performed better than the AV in this situation, Q3) In this situation, the AV performs well enough for me to engage in other activities, Q4) The situation was risky, Q5) The AV made an unsafe judgment in this situation, and Q6) The AV reacted appropriately to the environment. We measured the six STS-AD items with a 7-point Likert scale. This scale was used 21 times during the experiment, where the first measurement was administered after the first 10 videos to manipulate participants into a specific precondition (overtrust or undertrust) and the following 20 measurements were conducted immediately after each of the remaining 20 videos (see Fig. \ref{fig:procedure}). 
Behavioral measures were proven to be useful in measuring situational trust (e.g., \cite{hergeth2016}). Two questions were used to evaluate participants' behavioral trust adapted from \cite{yin2019}. The first question, “Would you like to take over control?”, was presented before the participants perceived how the AV handled the potential takeover transition (see Fig. \ref{fig:videosteps}). The aim of the question was to understand the point in time at which the participants would determine that taking over control was necessary. Thus, rather than the system-initiated takeover by issuing a TOR as shown in previous studies \cite{du2020psychophysiological,du2020predicting,zhou2020driver,Zhou:2019,du2019examining,zhou2021using}, the takeover was initiated based on participants' judgements on the safety of the situation by answering "Yes" to the first asked question \cite{wang2019exploring}. 
The second question, “Would you like to change your previous decision?”, was asked after the participant saw the vehicle's decision (see Fig. \ref{fig:videosteps}). With this order of questions, we evaluated participants' trust before and after viewing the vehicle's behavior. Overall behavioral trust was measured 30 times at the midpoint of each video (see Fig. \ref{fig:procedure}) because multiple measures were needed to calculate the agreement fraction and switch fractions (see Eq. \ref{agreement} and Eq. \ref{switch}) in the preconditions. To quantify the behavioral measure of trust, we modified the agreement and switch fractions suggested by Yin et al. \cite{yin2019}. The agreement fraction was based on participants’ willingness to take over control prior to seeing the vehicle's decision; however, the switch fraction was affected by participants’ willingness to change their previous takeover decision after seeing the vehicle's decision. Both measures are affected by trust. The agreement fraction is the number of scenarios for which participants' initial prediction agreed with the vehicle's decision divided by the total number of scenarios presented to the participants (see Eq. \ref{agreement}). The switch fraction is the number of scenarios for which participants' initial prediction disagreed with their final decision divided by the total number of scenarios (see Eq. \ref{switch}). 

\begin{equation}
A_k = \frac {\sum_{j=1}^{N} (P_{i_j}^k = D_j^k) }{N}, 
\label{agreement}
\end{equation}

\begin{equation}
S_k = \frac {\sum_{j=1}^{N} (P_{i_j}^k \neq P_{f_j}^k) }{N}, 
\label{switch}
\end{equation}
where $k$ refers to the $k$-th participant, $N$ is the total number of scenarios, $P_{i_j}^k$ refers to the $k$-th participant's initial prediction in $j$-th video, $P_{f_j}^k$ refers to the $k$-th participant final decision in $j$-th video, and $D_j^k$ refers to the vehicle's decision in $j$-th video for the $k$-th participant. In our study, the participants' initial prediction was based on their willingness to take over control prior to seeing the vehicle's decision. The vehicle's decision was whether the AV would fail or not in that particular scenario. The participants' final decision was based on their willingness to change their previous takeover decision after seeing the vehicle's decision.

\subsection{Survey Design and Procedure}
The experiment consisted of three sessions (corresponding to the three accuracy levels in a counterbalanced order) for one precondition, and each session was conducted with a two-day gap. The survey consisted of five sections, as illustrated in Fig. \ref{fig:procedure}. The first section included a consent form. In the second section, participants answered a set of demographic questions once during the first session. In the third section, participants were given a detailed explanation of the study procedure and went through a practice session. The explanations described the information provided in the video, such as surrounding vehicles and objects and the flow of actions. The explanations helped participants think like a passenger in order to make rational decisions. Participants then watched one training video that was different from the testing videos and answered the corresponding questions. In the fourth section, participants were required to watch 10 videos. If the tested precondition was overtrust, participants were required to watch 10 videos where the AV was able to safely handle the scenarios with no failures. However, if the tested precondition was undertrust, the 10 videos demonstrated AVs failing the driving scenarios. The order of the videos was randomized in the two conditions. In the fifth section, participants watched 20 videos of AV driving scenarios involving potential takeovers (corresponding to the participants' initiation of a takeover decision) with a different number of failures, depending on the tested AV accuracy (i.e., 1, 4, and 6 failures in 95\%, 80\%, and 70\% accuracy levels, respectively). For instance, if the tested AV accuracy was 95\%, one video showed that the AV was not able to handle the driving scenario successfully, while the remaining 19 videos showed successful driving scenarios with potential takeover transitions. The mean duration of each video was 26 s, and the time range value was from 23 s to 33 s. The order of the videos under each tested accuracy was randomized. The AV in the scenarios was not designed to take over control; however, we explained to the participants that the AV used in the videos represented an SAE Level 3 AV. During the study, each video was automatically paused for 9 s before showing the decision of the AV, and participants were asked whether they would like to take over control. The video automatically resumed to show the AV's decision. The participants were asked whether they would like to change their previous answer. Their decision should be based on the information provided about the surrounding vehicles and the previously watched videos (see Fig. \ref{fig:videosteps}). After each video (starting from the 10th video), the participants were required to answer the six questions on a 7-point Likert scale in the STS-AD survey. 


\subsection{Scenario Design}
\begin{figure} [bt!]
	\centering
	\subfloat[\label{fig:success}]{\includegraphics[width=.45\linewidth]{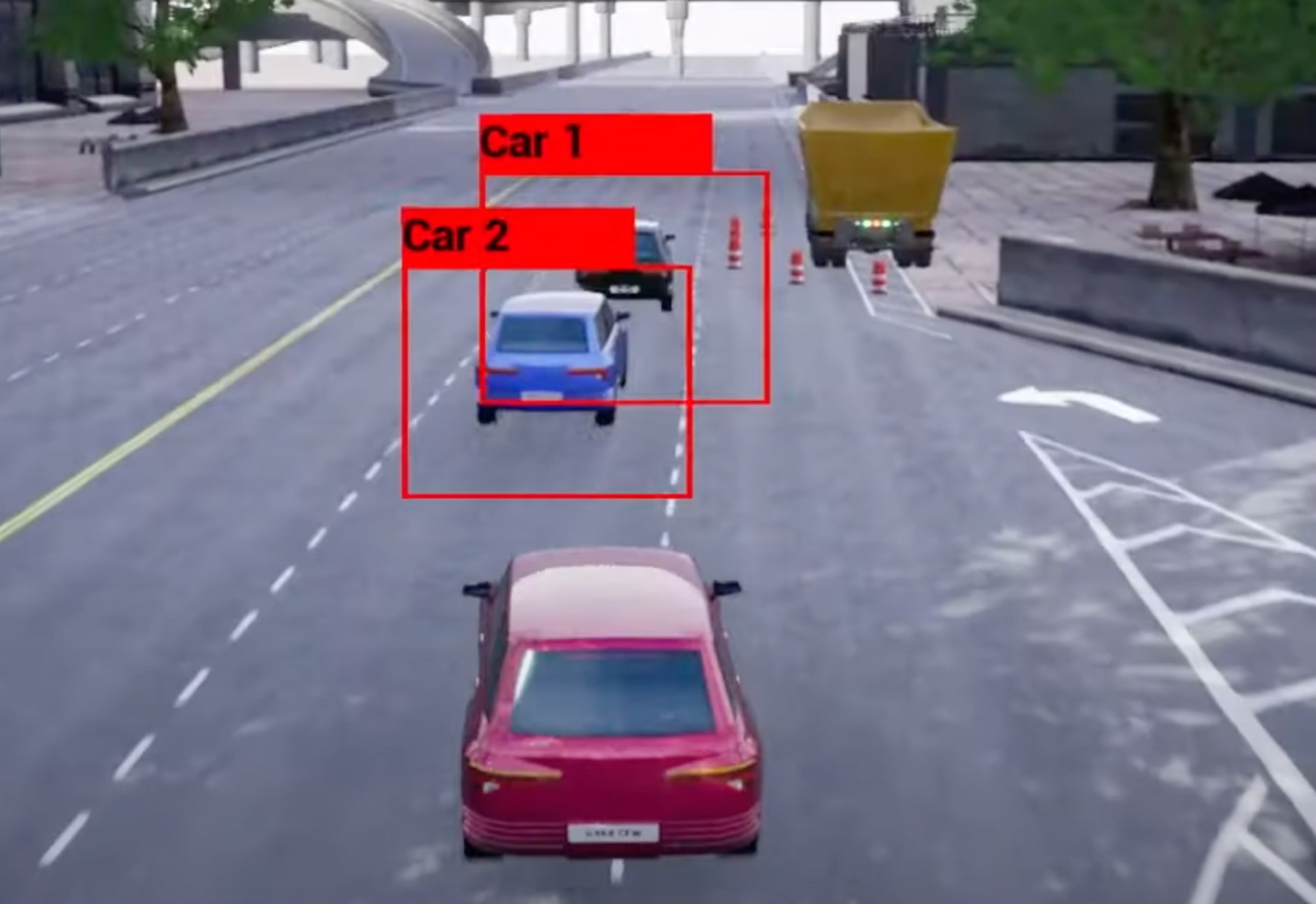}}
	\hspace{0pt}
	\subfloat[\label{fig:failure}]{\includegraphics[width=.44\linewidth]{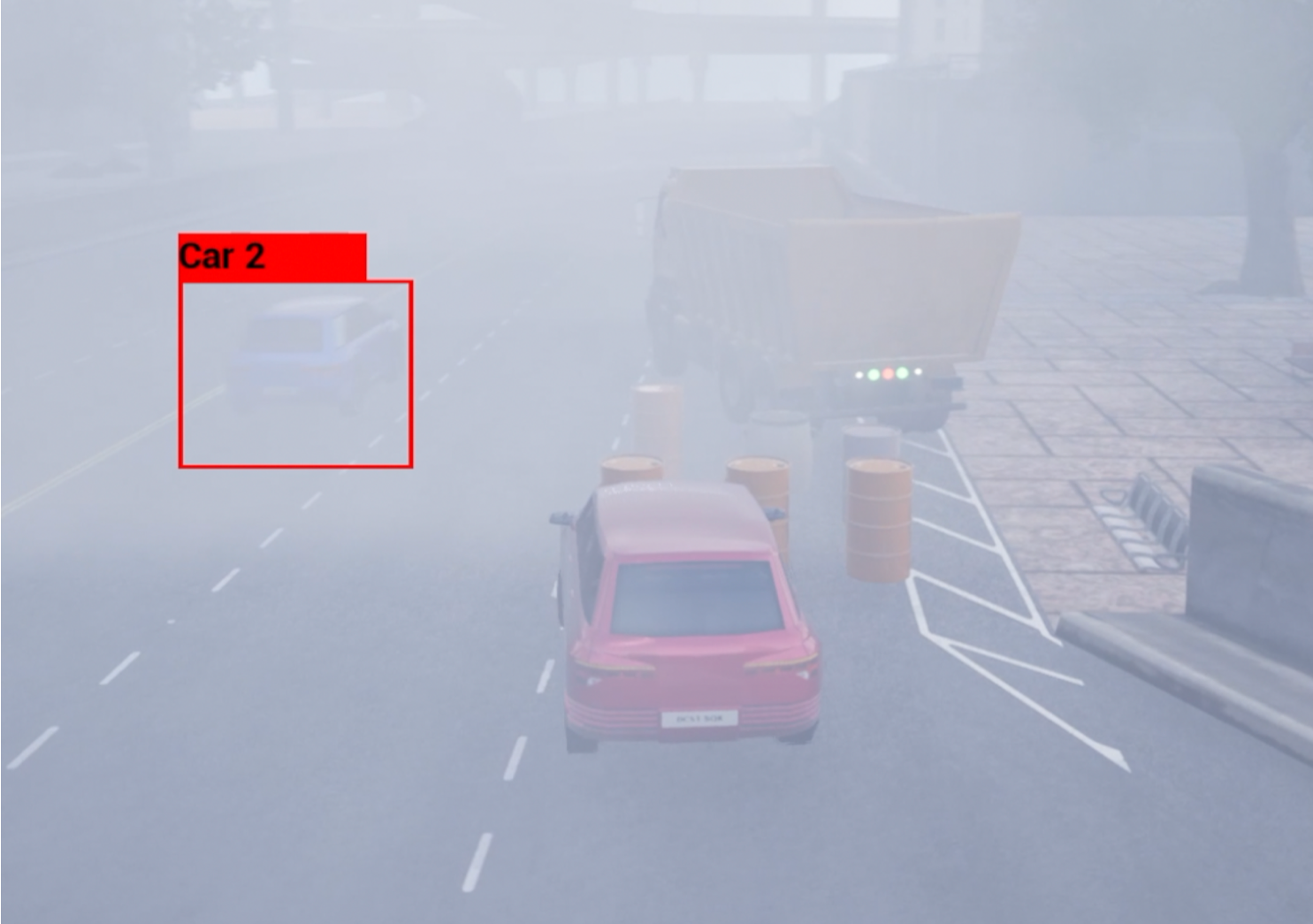}}
    \caption{Takeover scenario with (a) no failure and (b) failure}\label{fig:drivingScenarios}
\end{figure}

We investigated drivers' trust in AVs' takeover scenarios by exploring their situational trust in different system performances with different numbers of vehicle failures. We conducted an internet search to find takeover scenarios based on real experiences. From the web videos, we identified 30 different scenarios that we used to create the failure and nonfailure videos. In the failure scenarios, we added bad weather conditions, such as fog and snow, to improve the fidelity. Since adding an adverse weather condition to the scenarios might cue the participants and bias the results, bad weather (e.g., snow weather) was included in three of the nonfailure scenarios. In addition, in the training section, we made clear that AV failure could occur in both bad and good weather conditions depending on the scenarios. A brief description of the scenarios is shown in Table \ref{table:scenariodescription}. The failure scenarios were similar to the nonfailure scenarios except that the AV failed to handle the situation. For instance, one of the scenarios showed the AV traveling on a highway with a broken down truck parked at the side of the road. In the nonfailure scenario, the AV safely detected the truck and changed lanes before hitting the truck (see Fig. \ref{fig:success}). However, in the failure scenario, the AV failed to detect the truck due to foggy conditions and crashed into the truck (see Fig. \ref{fig:failure}). 

\begin{figure*}[h!]
\centering
\includegraphics[width=0.8\linewidth]{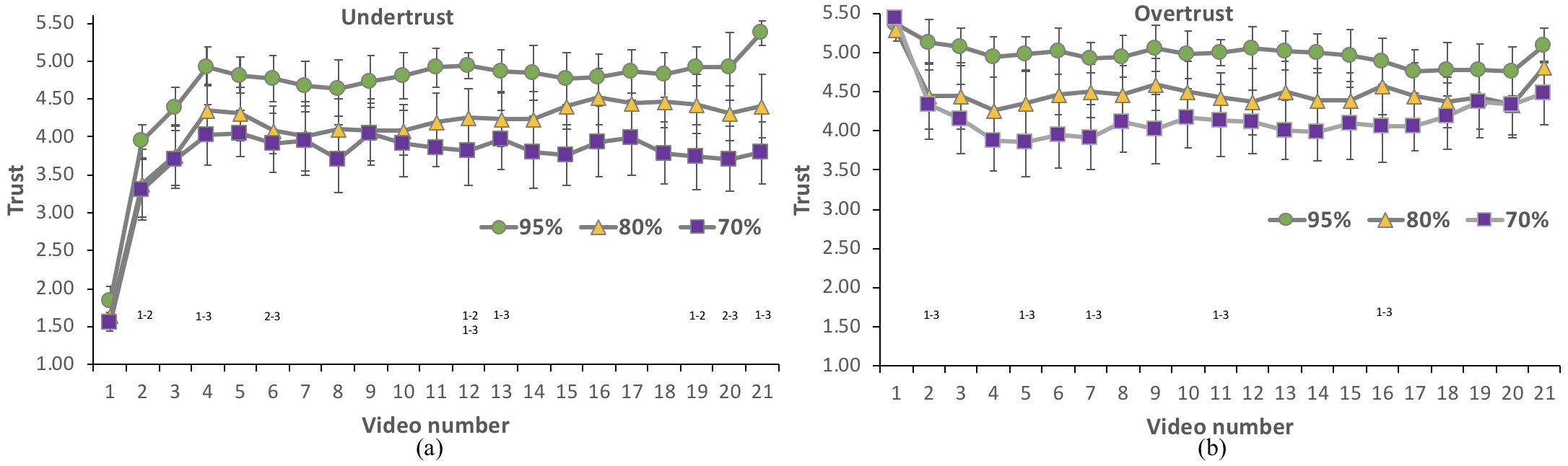}
\caption{Overall mean and standard error of self-reported situational trust (SST) measured by the STS-AD six scales for all the participants at different accuracy levels and trust preconditions. (a) Undertrust precondition. (b) Overtrust precondition. “1” = strongly distrust and “7” = Strongly trust. Note that the first video in the graph represents the average results of the first 10 videos that the participants watched to manipulate them into an undertrust or overtrust condition. Along the x-axis, the accuracy levels that have significant differences of pairwise comparisons are indicated with number pairs. ``1" indicates 95\%, ``2" indicates 80\%, and ``3" indicates 70\%. For example, in the second video in (a), there was a significant difference between 95\% and 80\%.}
\label{fig:overall}
\end{figure*}

\subsection{Data Analysis}
Statistical analysis was conducted using SPSS statistics software (IBM, NY). An analysis of variance (ANOVA) was used to analyze the effects of system performance and trust preconditions on dynamic situational trust. The alpha level was set at 0.05 for all the statistical tests. The ANOVA  assumptions, including normality and homogeneity of variance, were not violated for either overtrust or undertrust preconditions. Pairwise comparisons were performed with Bonferroni correction.

\section{RESULTS}
\subsection{Manipulation Check}
We compared the self-reported situational trust (SST) in the precondition stage (i.e., the manipulation stage with 10 videos of consecutive failures or successes, see Fig. \ref{fig:procedure}) with that before the manipulation and that after the manipulation. Before the manipulation stage, a question (i.e., in general, how much do you trust an AV?) was asked in the demographic section of the survey. 
In the undertrust precondition, we found that participants' SST (see the 1st video in Fig. \ref{fig:overall}a) was significantly lower ($F(1,20)=128.337, p = .000$) than the overall trust level before manipulation and was significantly lower than that in the testing stage by aggregating the last 20 videos in the 95\% ($p = .000$), 80\% ($p = .000$), and 70\% ($p = .000$) accuracy levels. 

In the overtrust precondition, we found that participants' SST ($M=5.327, S.E.= .113$, see the 1st video in Fig. \ref{fig:overall}b) was significantly higher ($F(1,20)= 12.910, p = .002$) than the overall trust level before manipulation and was significantly higher than in the testing stage by aggregating the last 20 videos in the 95\% ($p = .001$), 80\% ($p = .000$), and 70\% ($p = .000$) accuracy levels. Furthermore, in the undertrust precondition, participants' self-reported situational trust (SST) did not have significant differences ($F(2,40)= 1.709, p = .194$) among the three accuracy conditions. Similarly, for the overtrust precondition (see Fig. \ref{fig:overall}b), participants' SST had no significant differences ($F(2,40)= .342, p = .712$) among the three accuracy levels. These results showed that participants' trust was calibrated to overtrust and undertrust the AV.

For the agreement fraction, in the manipulated precondition (i.e., in the first 10 videos), using a two-way mixed ANOVA, we did not find any significant differences among the three accuracy levels ($F(2,80)=1.429, p = .246$), two preconditions ($F(1,40)=.041, p = .218$), or interaction between accuracy and preconditions ($F(2,80) = 1.554, p = .842$). We further compared the agreement fraction between the first 10 videos and the last 20 videos, and no significant differences were found. For the switch fraction, using a two-way mixed ANOVA, we did not find any significant differences among the three accuracy levels ($F(2,80)= 2.669, p = .076$) or interaction between accuracy and preconditions ($F(2,80)= 1.836, p = .167$). However, we found that those in the undertrust precondition had a significantly higher switch fraction than those in the overtrust precondition ($F(1,40)= 9.484, p = .004$). We further compared the switch fraction between the first 10 videos and the last 20 videos, and we did not find any significant differences.

\subsection{Self-reported Situational Trust}

Due to a relatively small number of participants, we calculated the mean of the SST over all 21 participants from the second video (in Fig. \ref{fig:overall}, i.e., the 11th video in the experiment) with a smoothing function by using a window length of 5 videos (see Eq. \ref{smooth}). Note that the first video in the figure represents the average results of the first 10 videos that the participants watched to manipulate them in an undertrust or overtrust condition, which was not smoothed with the other following videos.
\begin{equation}
T_s(i) = \frac {1}{2N+1} (T(i+N) + T(i+N-1)+...+ T(i-N)),
\label{smooth}
\end{equation}
where $T_s(i)$ is the smoothed value of SST, $N=2$ is the number of neighboring data points on both sides of, $T_s(i)$, and $2N+1=5$ is the size of the sliding window. Note that the smoothing function was used only to show the dynamic changes of situational trust in the figures, while this was not applied to the statistical data analysis below.


\begin{figure}[bt!]
\centering
\includegraphics[width=0.9\linewidth]{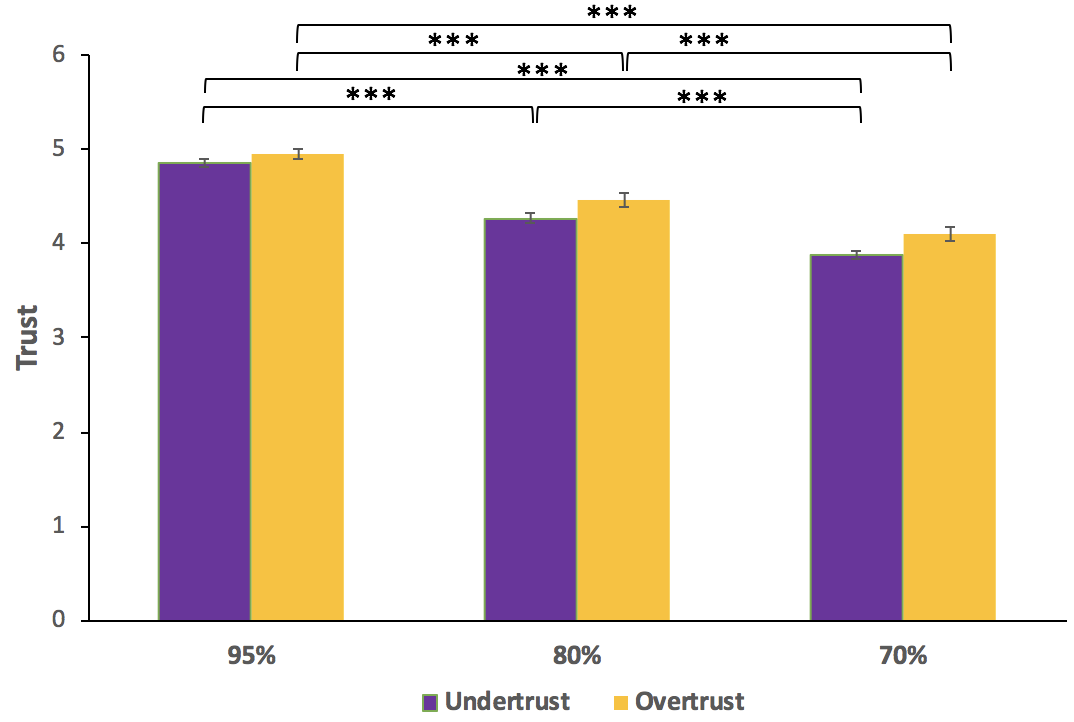}\hfill
\caption{Mean self-reported situational trust at different accuracy and trust precondition levels with standard errors, where `*' indicates $p<0.05$,`**' indicates $p<0.01$, and `***' indicates $p<0.001$.}
\label{fig:overall1}
\end{figure}

Fig. \ref{fig:overall1} illustrates the mean SST for the three tested accuracy levels under the two trust preconditions (i.e., overtrust and undertrust). The first video in Fig. \ref{fig:overall} (i.e., average of the 10 preconditioning videos) was removed before conducting the following statistical analysis. A mixed two-way ANOVA showed that the main effect of accuracy was significant ($F(2,80)=144.794, p =.000$) whereas the main effect of trust preconditions was not significant ($F(1,40)= .915, p = .344$) (see Fig. \ref{fig:overall1}). There was no significant interaction effect between the trust preconditions and the tested accuracy levels ($F(2,80)= 0.269, p = .765$). The pairwise comparison showed a significant difference between 95\% and 80\% accuracies ($p = .000$), 95\% and 70\% accuracies ($p = .000$), and 80\% and 70\% accuracies ($p = .000$) (see Fig. \ref{fig:overall1}). Due to the significant main effect of accuracy levels, we also conducted a pairwise comparison among three accuracy levels at each video as illustrated along the x-axis in Fig. \ref{fig:overall} for both undertrust and overtrust preconditions. Significant differences were labeled by the numbers in the figure, where ``1" indicates 95\%, ``2" indicates 80\%, and ``3" indicates 70\%.

\begin{figure*}[h!]
\centering
\includegraphics[width=14cm]{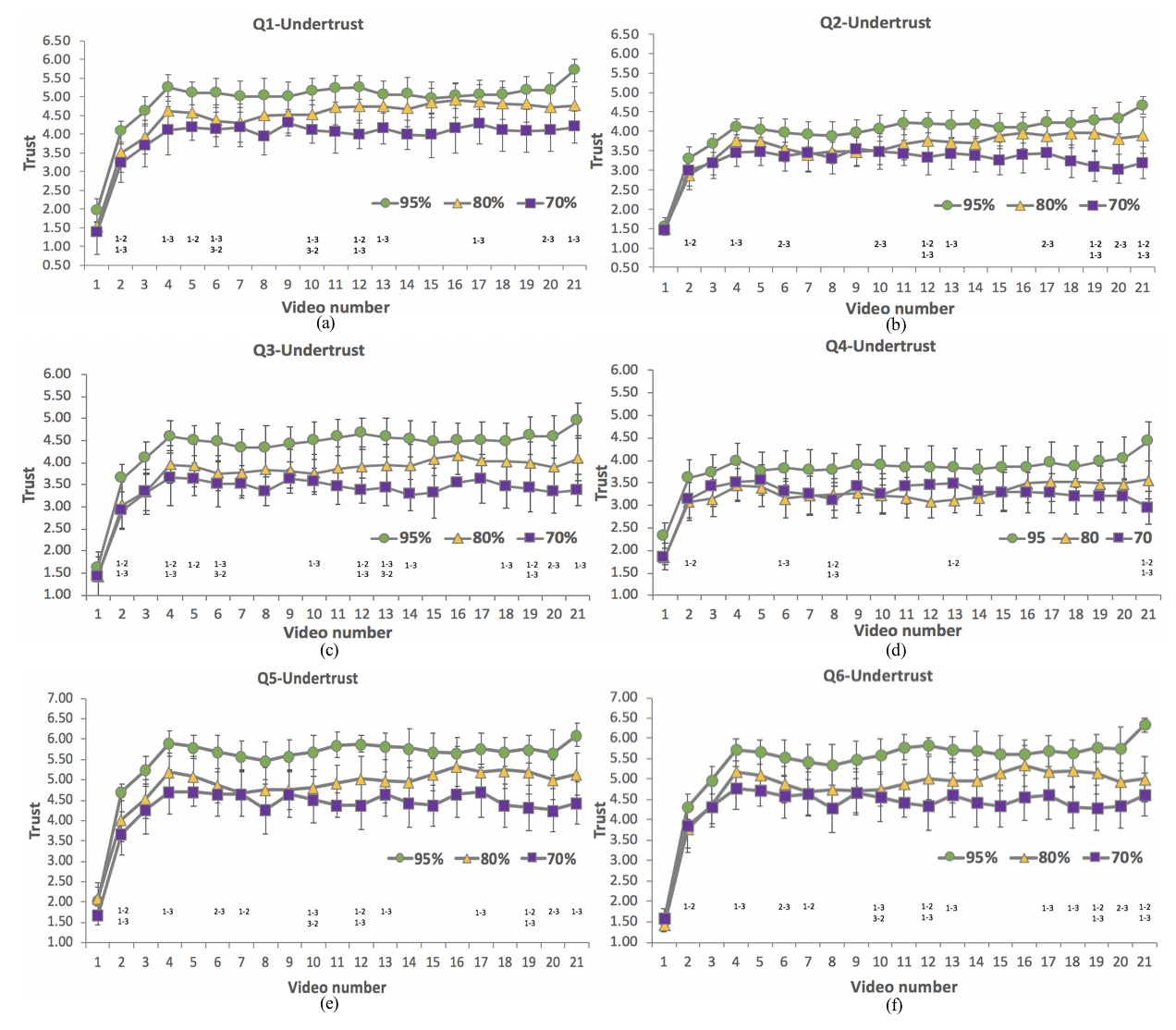}
\caption{Mean measures of the STS-AD six scales (see dependent variables in Section III Method) for all the participants in the undertrust precondition and at different accuracy levels. Note that the first video in the graph represents the average results of the first 10 videos that the participants watched to manipulate them into an undertrust condition. Along the x-axis, the accuracy levels having a significant difference in pairwise comparisons are indicated with number pairs. ``1" indicates 95\%, ``2" indicates 80\%, and ``3" indicates 70\%. For example, in the second video in Q1, there was a significant difference between 95\% and 80\% and between 95\% and 70\%. The error bars indicate the standard errors of the means.}
\label{fig:STS-ADUndertrust}
\end{figure*}

\begin{figure*}[h!]
\centering
\includegraphics[width=14cm]{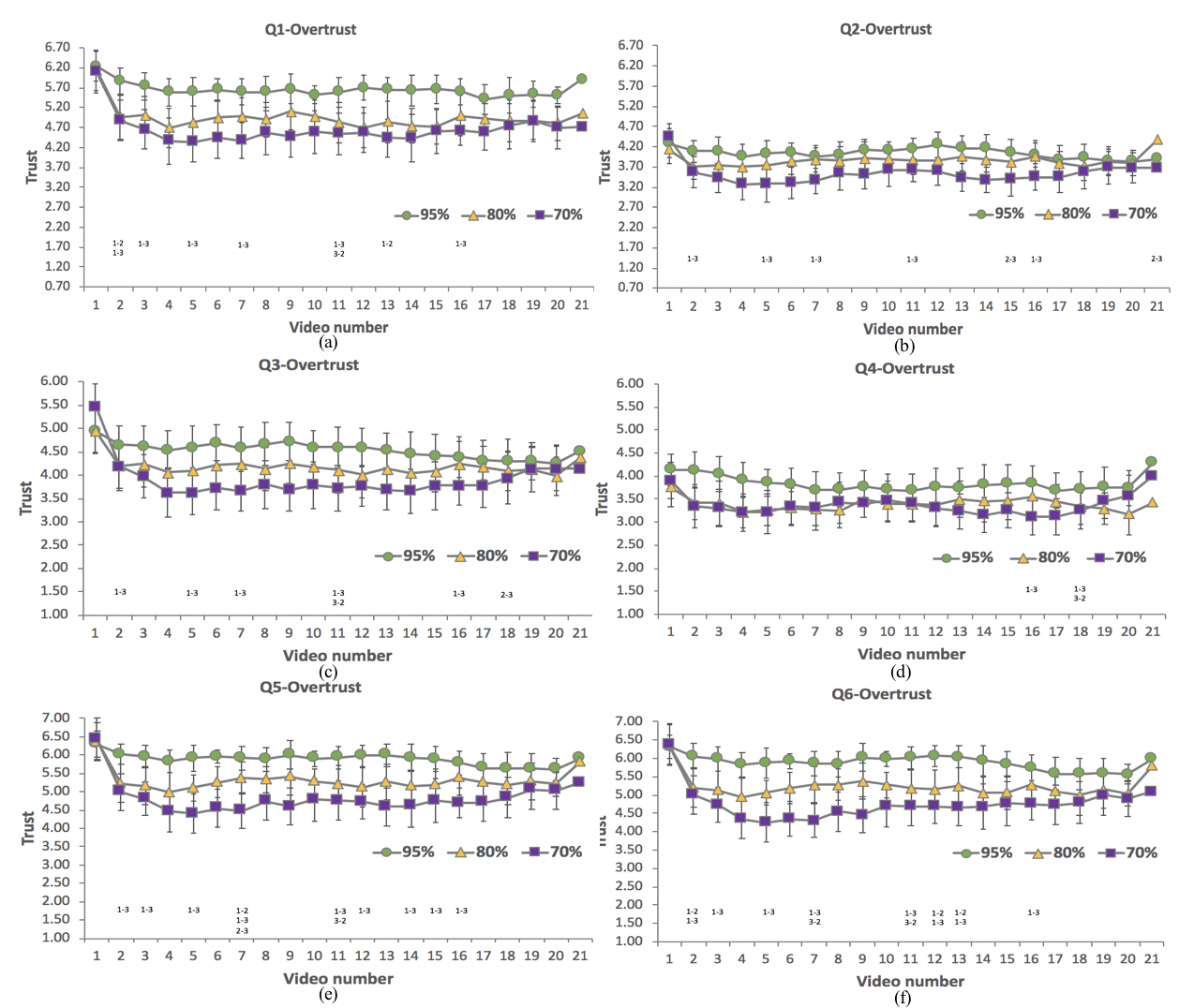}
\caption{Mean measures of the STS-AD six scales (see dependent variables in Section III Method) for all participants in the overtrust precondition and at different accuracy levels. Note that the first video in the graph represents the average results of the first 10 videos that the participants watched in order to manipulate them in an overtrust condition. Along the x-axis, the accuracy levels having a significant difference in pairwise comparisons are indicated with number pairs. ``1" indicates 95\%, ``2" indicates 80\%, and ``3" indicates 70\%. For example, in the second video in Q1, there was a significant difference between 95\% and 80\% and between 95\% and 70\%. The error bars indicate the standard errors of the means.}
\label{fig:STS-ADOvertrust}
\end{figure*}
TORs in conditional AVs can be interpreted as a system failure since the AV is not able to handle the situations
Furthermore, we analyzed each of the STS-AD trust scales separately, as shown in Fig. \ref{fig:STS-ADUndertrust} and Fig. \ref{fig:STS-ADOvertrust}, using a two-way ANOVA to identify sensitive items since the experiment was conducted in low fidelity. For instance, the risk item can be perceived as lower than in real situations. We found significant main effects of accuracy levels for all the six questions of the STS-AD scale (all $p = .000$). In addition, we conducted a pairwise comparison at each video, as illustrated along the x-axis in Fig. \ref{fig:STS-ADUndertrust} and Fig. \ref{fig:STS-ADOvertrust}. Whenever a significant difference existed at each video, they were labeled by the numbers in the figure. There was a significant main effect for trust precondition for Q1 ($F(1,40)= 4.293, p = .045$). Pairwise comparison showed significant differences when comparing 95\% vs. 95\% ($p = .000$), 80\% vs. 80\% ($p = .007$), and 70\% vs. 70\% ($p = .000$) between two preconditions. For Q2, Q3, Q4, and Q6, there was no significant difference between the two trust conditions. For Q5, there was a significant main effect for trust precondition ($F(1,40)= 4.530, p = .040$), and significant differences were found when comparing 80\% vs. 80\% ($p = .019$) and 70\% vs. 70\% ($p = .009$) between the two preconditions.

\begin{figure*}[h!]
\centering
\includegraphics[width=1\linewidth]{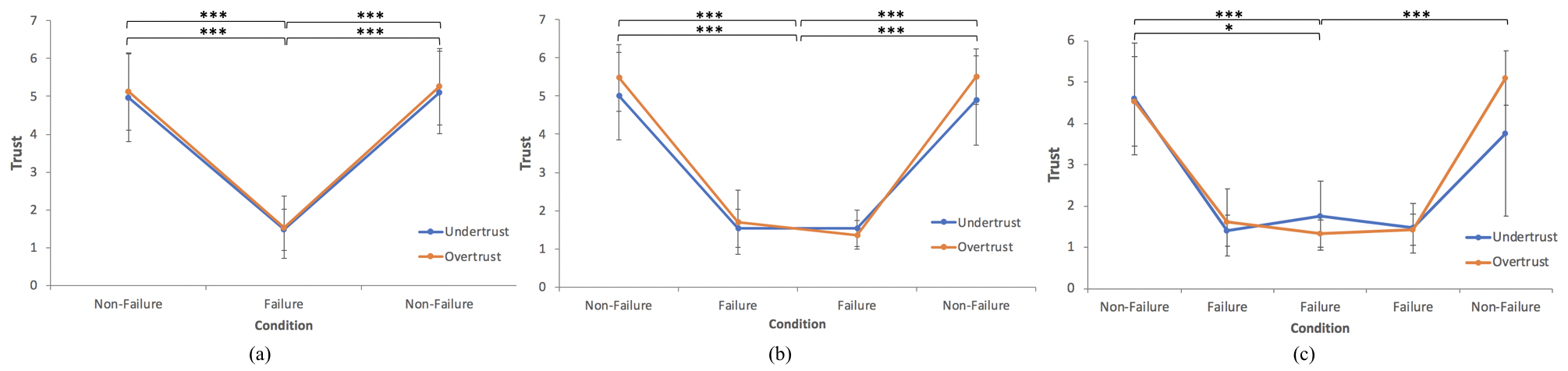}
\caption{Mean measure of the STS-AD six scales in the overtrust and undertrust precondition at different consecutive failure occurrences with standard deviation, where `*' indicates $p<0.05$,`**' indicates $p<0.01$, and `***' indicates $p<0.001$. (a) 1 failure. (b) 2 failures. (c) 3 failures.}
\label{consecFailure}
\end{figure*}

We grouped the results according to the transition condition of the videos. We identified four patterns of failure occurrences, including one failure (i.e., 113 occurrences in the undertrust precondition and 123 occurrences in the overtrust precondition), two consecutive failures (i.e., 31 occurrences in the undertrust precondition and 27 occurrences in the overtrust precondition), three consecutive failures (i.e., 5 occurrences in the undertrust precondition and 8 occurrences in the overtrust precondition), and four consecutive failures (i.e., 2 occurrences in the undertrust precondition and 2 occurrences in the overtrust precondition). The situations where the failure scenario occurred as the first or last video in the 20 video sequence were removed from this analysis. The case with four consecutive failures was not analyzed since it only occurred two times in the undertrust and overtrust preconditions.
In the case of one failure video, there was a significant difference in the SST level between the failure video and the previous nonfailure video (i.e., undertrust $F(1,112)= 817.466, p = .000$; overtrust $F(1, 122)= 754.016, p = .000)$) and between the failure video and the following nonfailure video (i.e., undertrust $F(1, 112)= 968.681, p = .000$; overtrust $F(1, 122)= 662.665, p = .000)$ (see Fig. \ref{consecFailure}a). In the case of two consecutive failure videos, there was a significant difference in the SST level between the average of the two consecutive failure videos and the previous non-failure video (i.e., undertrust $F(1, 30)= 253.144, p = .000$; overtrust $F(1, 26)= 339.807, p = .000$) and between the average of the two consecutive failure videos and the following nonfailure video (i.e., undertrust $F(1, 30)= 267.374, p = .000$; overtrust $F(1, 26)= 405.087, p = .000$) (see Fig. \ref{consecFailure}b). In the case of three consecutive failure videos, there was a significant difference in the SST level between the average of the three consecutive failure videos and the previous non-failure video (i.e., undertrust $F(1, 4)= 19.033, p = .012$; overtrust $F(1, 7)= 49.860, p = .000$) and between the average of the three consecutive failure videos and the following nonfailure video in the overtrust precondition (i.e., $F(1, 14)= 113.370, p = .000$). There was a marginally significant difference between the average of the three consecutive failure videos and the following nonfailure video in the undertrust precondition (i.e., $F(1, 4)= 6.459, p = .064$) (see Fig. \ref{consecFailure}c).

\subsection{Behavioral Situational Trust}
\begin{figure}[h!]
\centering
\includegraphics[width=0.8\linewidth]{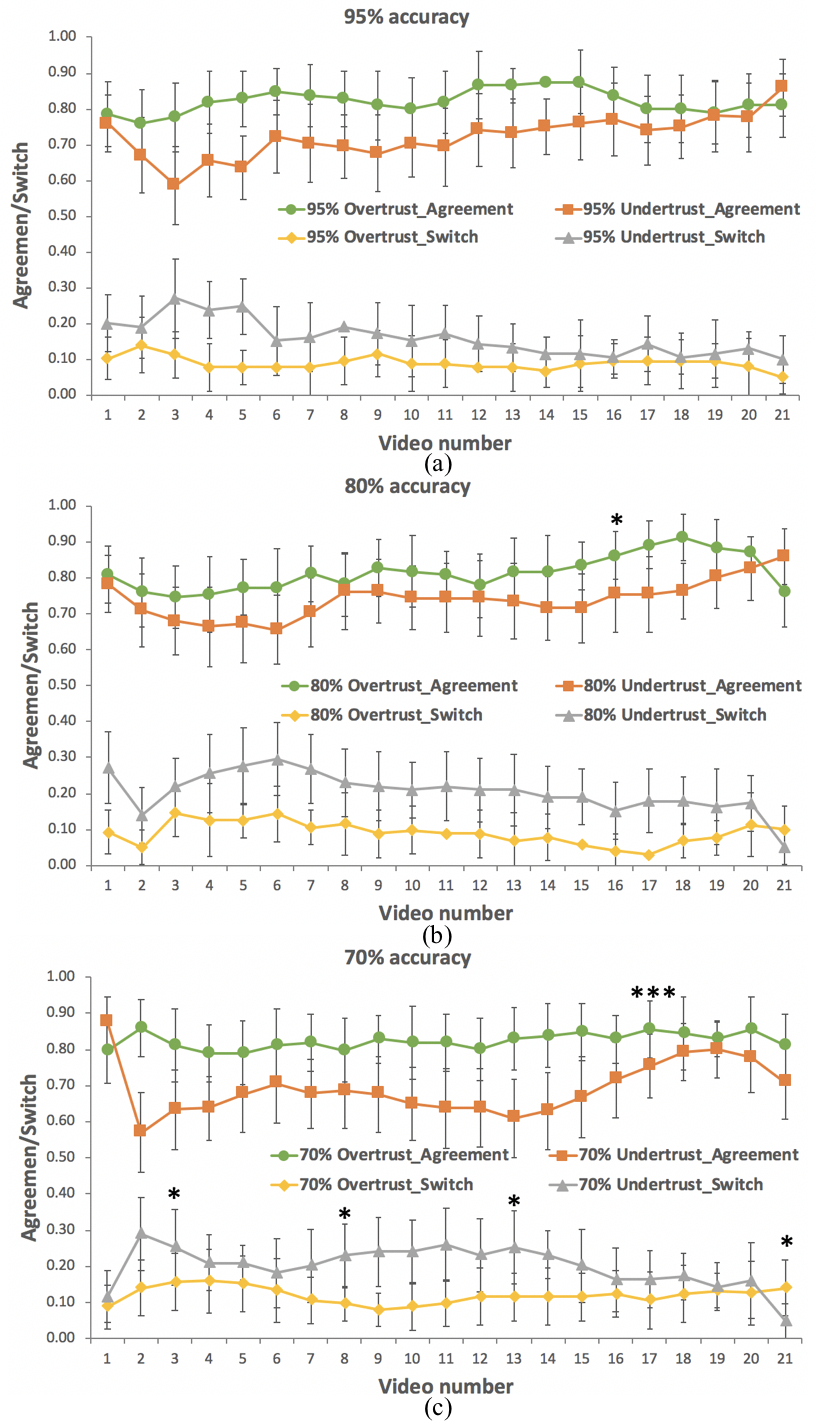}
\caption{Overall mean and standard error of behavioral situational trust measured by the agreement and switch fractions for all the participants at different accuracy levels and trust preconditions. (a) 95\% accuracy. (b) 85\% accuracy. (c) 70\% accuracy. Note that the first video in the graph represents the average results of the first 10 videos that the participants watched to manipulate them in an undertrust or overtrust condition, where `*' indicates $p<0.05$,`**' indicates $p<0.01$, and `***' indicates $p<0.001$.}
\label{fig:70A}
\end{figure}

\begin{figure}[h!]
\centering
\includegraphics[width=0.8\linewidth]{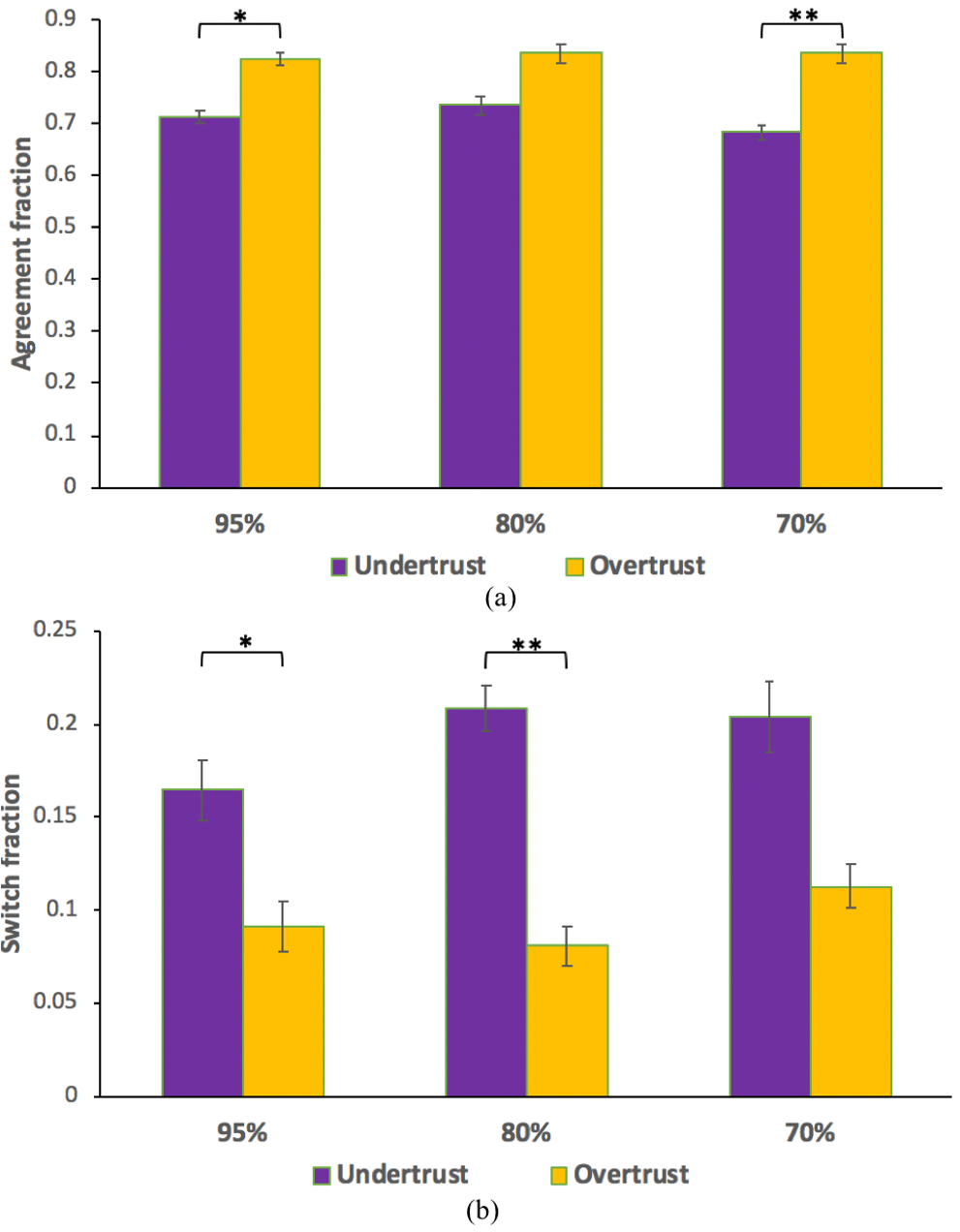}
\caption{BST: (a) Mean agreement fraction with standard error and (b) Mean switch fraction with standard error at different accuracy levels and in the overtrust and undertrust preconditions, where `*' indicates $p<0.05$,`**' indicates $p<0.01$, and `***' indicates $p<0.001$.}
\label{fig:STS-AgreeSwitch}
\end{figure}

As shown in Fig. \ref{fig:STS-AgreeSwitch}a, for the agreement fraction, the main effect of accuracy was not significant on BST ($F(2,80)= .497, p = .61$), whereas the main effect of precondition was significant ($F(1,40)= 8.553, p = .006$). There was no significant interaction effect between the trust precondition and the tested accuracy levels ($F(2,80)= .483 , p = .619$). Pairwise comparison showed that the agreement fraction was significantly higher in the overtrust condition than in the undertrust condition when comparing 95\% vs. 95\% ($p = .029$) and 70\% vs. 70\% ($p = .006$) and marginally higher in the overtrust condition than that in the undertrust condition when comparing 80\% vs. 80\% ($p = .087$). 

For the switch fraction as shown in Fig. \ref{fig:STS-AgreeSwitch}b, the main effect of accuracy levels was not significant on BST ($F(2,80)= 1.888, p = .158$), whereas the main effect of precondition was significant ($F(1,40)= 10.053, p = .003$). There was no significant interaction effect between the trust precondition and the tested accuracy levels ($F(2,80)= 0.657, p = .521$). Pairwise comparison showed that the switch fraction in the undertrust condition was significantly higher than in the overtrust condition when comparing 95\% vs. 95\% ($p = .023$) and 80\% vs. 80\% ($p = .004$), and on the borderline of significance when comparing 70\% vs. 70\% ($p = .050$) between the two preconditions. 
Due to the significant main effect of preconditions, we also conducted a pairwise comparison at each video, as illustrated in Fig. \ref{fig:70A}, for both undertrust and overtrust preconditions for the agreement and switch fractions. 

\section{Discussions}

\subsection{Self-reported Situational Trust}
In this study, we aimed to understand the effects of trust preconditions and system performance on dynamic situational trust in conditional AVs. First, by letting participants watch 10 videos of takeover scenarios with or without failures, we were able to manipulate them into an overtrust or undertrust condition by examining the SST measure. However, we were not able to measure BST before the experiment, so we were not able to test BST for a manipulation check with respect to a pretest comparison. We were not able to find any significant differences between the preconditions and the test conditions. This indicated that the BST was not good at measuring trust levels manipulated by vehicle performance levels, which was also evidenced by the insignificant main effect of accuracy levels.

The SST measure was quickly calibrated to the different accuracy levels from their corresponding trust preconditions (see Fig. \ref{fig:overall}). We noticed that participants' average SST level between the $10th$ and the $11th$ video increased from 1.563 to 3.300 for 70\% ($p = .000$), from 1.643 to 3.383 for 80\% ($p = .000$), and from 1.841 to 3.947 for 95\% ($p = .000$) accuracy levels (see Fig. \ref{fig:overall}a). Whereas in Fig. \ref{fig:overall}b, we noticed that participants' average SST level at the $10th$ video decreased significantly from 5.294 to 4.450 for 80\% ($p = .010$) and from 5.452 to 4.333 for 70\% ($p = .001$) accuracy levels. For 95\% accuracy, there was no significant difference ($p = .341$) in trust between the $10th$ and the $11th$ video. It showed that the participants were able to quickly calibrate their SST based on the performance of the AV. This effect seemed to be more prominent in the undertrust precondition than in the overtrust precondition by comparing Fig. \ref{fig:overall}a and Fig. \ref{fig:overall}b for the first several SST scores. In the undertrust precondition, participants took more time (i.e., until the $4th$ video after the precondition) to calibrate their trust after watching the 10 failure videos. However, in the overtrust precondition, the participant calibrated their trust faster (i.e., until around the $2nd$ video after the precondition). This seemed to be consistent with previous findings \cite{parasuraman2010complacency} that failures led to a steep decrease in trust and that the recovery was slow. We noticed an increase in trust for the last video in the undertrust and the overtrust preconditions at the three tested accuracies (see Fig. \ref{fig:overall}). Due to the random order assigned in Qualtrics, the majority of the scenarios for the last video were nonfailure scenarios.  

Second, our findings showed that the participants were able to perceive the system performance at different accuracy levels and to adjust their SST based on the system performance (see Fig. \ref{fig:STS-ADUndertrust} and Fig. \ref{fig:STS-ADOvertrust}), even though the accuracy of the system was not presented to the participants. We noticed that regardless of the trust precondition of the participants, they always had a higher SST at 95\% compared to 80\% and 70\%. This is consistent with previous studies (e.g., \cite{hergeth2016,beggiato2013}) finding that participants were capable of learning the performance of the AV dynamically and calibrating their SST level over time by understanding the capabilities and limitations of the AV over time. This also indicated that the SST levels closely reflected the actual performance of the AVs, which could help avoid misuse and abuse of conditional AVs.

\subsection{Behavioral Situational Trust}

Agreement and switch fractions were used in this study as BST measures of participants' dynamic situational trust. First, in Fig. \ref{fig:STS-AgreeSwitch}a, we noticed that the agreement fraction in the overtrust precondition was significantly higher than in the undertrust precondition for all the tested accuracy levels. In Fig. \ref{fig:STS-AgreeSwitch}b, the switch fraction was significantly higher in the undertrust condition than that in the overtrust precondition. However, there was no significant main effect for different accuracy levels. Hence, the SST and BST were complementary in that the SST measure was sensitive to system performance, while the BST was sensitive to trust preconditions. The agreement fraction indicated participants' reliance on the AV to a large extent (automation reliance), while the switch fraction indicated participants' rejection of the AV's decisions (self-reliance) \cite{dzindolet2003role}. When the participants viewed the first 10 videos with all successes, they tended to overtrust the vehicle and believed that it was at least as reliable as, if not more reliable than, manual operations. This made them more likely to rely on the automation and agree with the system's decisions. In  contrast, when participants were in the undertrust condition, they tended to have a low consistency with the system's decisions, which made them switch their initial prediction about whether the system would handle the driving scenarios. Such results were supported by previous findings that a high level of trust was associated with more reliance on automation and vice versa \cite{wickens2015engineering}. The difficulty in reducing such automation bias (e.g., automation reliance and self-reliance) might also explain why different levels of system performance did not play a role in BST when participants were in an overtrust or undertrust preconditions.

\subsection{Comparison between SST and BST}

The results showed that the SST and BST were complementary to each other. However, Yin et al. \cite{yin2019} showed a similar trust pattern in the analysis of self-reported and behavioral measures. Their self-reported and behavioral measures were different from those used in our work. Murtin et al. \cite{murtin2018trust} showed that both self-reported and behavioral measures are correlated with expected trustworthiness, but behavioral trust additionally captures the willingness to cooperate during a specific interaction. They concluded that these two measures are related but should be considered as complementary. Our results showed that the SST measure was only sensitive to system performance since the main effect of accuracy was significant. By analyzing the items of the STS-AD scale, we noticed that only Q1 (i.e., I trust the automation in this situation) and Q5 (i.e., The AV made unsafe judgment in this situation) had a significant main effect of trust precondition. This result might be caused by the low fidelity of the system, which made it difficult to estimate the performance, NDRT, risk, and reaction to the environment. Another reason might be that the participants took a short period of time to develop similar SST levels between the overtrust and undertrust preconditions. Furthermore, our results showed that the BST was sensitive only to trust preconditions since the main effect of precondition was significant. 
In our experiment, the SST and BST were not measured simultaneously,
which could explain why the reason for not obtaining the same trust pattern between these two measures was not attained. BST was measured during the scenario, while SST was measured after the scenario was completed.



\subsection{Implications}
Trust plays an important role in the adoption and proper use of AVs, and different constructs of trust could have different features. Situational trust evolves depending on multiple factors in the human-AV interaction process. Our study showed that the participants were able to calibrate their SST to the real performance of the AV over time. This indicates that clearly demonstrating the capabilities and limitations of the AV can help drivers quickly calibrate their trust level \cite{lee2004}. However, the calibrated trust level was not influenced by the precondition of the participants'. The effect of trust preconditions was reflected by the BST measures, which could be considered as learned trust in AVs in this study. If such learned trust is inconsistent with the system performance, automation bias can occur. Thus, it is important to consider drivers' learned trust in AVs when designing calibration systems, as these preconditions could potentially influence their calibrated situational trust over time. We also demonstrated that two types of situational trust measures, i.e., SST and BST, were complementary, and the inconsistency between them calls for further investigation into the reliability of different measures.

\section{Conclusions and Future Research}

In this work, we investigated the effects of system performance and participants' trust preconditions on dynamic situational trust in conditional AVs. Dynamic situational trust was measured using self-reported and behavioral measures, and participants were able to adjust their SST levels dynamically to be consistent with the performance of the AV. However, such results were moderated by participant trust preconditions, as measured by BST levels. Such insights revealed important implications for designing a calibration system for conditional AVs.

Our study also has limitations, which can be left for future research. First, this study was conducted in a low-fidelity experimental setup with a small sample size. To determine whether there are consistent results, future studies should be conducted in a high-fidelity driving simulator or even in a naturalistic driving environment with a larger and diverse sample size. Second, trust was mainly evaluated in takeover scenarios in conditional AVs using SST with the STS-AD scales and BST with the agreement and switch fractions. Thus, further analyses are needed to explore trust in other types of scenarios with other possible measures, such as eye-tracking data \cite{hergeth2016}. Third, failure scenarios occured only in poor weather conditions, which might cause participants' to be biased during the trust evaluation. Additionally, for the first 10 videos, the average SST was calculated only once at the end of the 10th video to save time. In the undertrust precondition, after watching 10 failure videos, participants might doubt that the AV was not capable. The participants' knowledge level regarding the AV capabilities could play a role in their trust formation, which was not studied in this paper. Fourth, this study mainly investigated the effects of trust preconditions and system performance on dynamic situational trust. Future studies could potentially include other factors, such as cognitive workload, and explore the two other layers of trust (e.g., learned trust and dispositional trust) to gain a complete understanding of the effects of system performance and trust precondition.

\bibliography{main}

\begin{thebibliography}{10}
\providecommand{\url}[1]{#1}
\csname url@samestyle\endcsname
\providecommand{\newblock}{\relax}
\providecommand{\bibinfo}[2]{#2}
\providecommand{\BIBentrySTDinterwordspacing}{\spaceskip=0pt\relax}
\providecommand{\BIBentryALTinterwordstretchfactor}{4}
\providecommand{\BIBentryALTinterwordspacing}{\spaceskip=\fontdimen2\font plus
\BIBentryALTinterwordstretchfactor\fontdimen3\font minus
  \fontdimen4\font\relax}
\providecommand{\BIBforeignlanguage}[2]{{%
\expandafter\ifx\csname l@#1\endcsname\relax
\typeout{** WARNING: IEEEtran.bst: No hyphenation pattern has been}%
\typeout{** loaded for the language `#1'. Using the pattern for}%
\typeout{** the default language instead.}%
\else
\language=\csname l@#1\endcsname
\fi
#2}}
\providecommand{\BIBdecl}{\relax}
\BIBdecl

\bibitem{xu2017}
Z.~Xu, M.~Wang, F.~Zhang, S.~Jin, J.~Zhang, and X.~Zhao, ``Patavtt: A
  hardware-in-the-loop scaled platform for testing autonomous vehicle
  trajectory tracking,'' \emph{Journal of Advanced Transportation}, vol. 2017,
  2017.

\bibitem{rice2019}
D.~Rice, ``The driverless car and the legal system: Hopes and fears as the
  courts, regulatory agencies, waymo, tesla, and uber deal with this exciting
  and terrifying new technology,'' \emph{Journal of Strategic Innovation and
  Sustainability}, vol.~14, no.~1, pp. 134--146, 2019.

\bibitem{kohli2019}
P.~Kohli and A.~Chadha, ``Enabling pedestrian safety using computer vision
  techniques: A case study of the 2018 uber inc. self-driving car crash,'' in
  \emph{Future of Information and Communication Conference}.\hskip 1em plus
  0.5em minus 0.4em\relax Springer, 2019, pp. 261--279.

\bibitem{zhang2019determinants}
B.~Zhang, J.~de~Winter, S.~Varotto, R.~Happee, and M.~Martens, ``Determinants
  of take-over time from automated driving: A meta-analysis of 129 studies,''
  \emph{Transportation research part F: traffic psychology and behaviour},
  vol.~64, pp. 285--307, 2019.

\bibitem{cao2021towards}
Y.~Cao, F.~Zhou, E.~Pulver, L.~Molnar, L.~Robert, D.~Tilbury, X.~J. Yang
  \emph{et~al.}, ``Towards standardized metrics for measuring takeover
  performance in conditionally automated driving: A systematic review,'' 2021.

\bibitem{edmonds2019}
E.~Edmonds, ``Three in four americans remain afraid of fully self-driving
  vehicles,'' \emph{American Automobile Association, https://newsroom. aaa.
  com/2019/03/americansfear-self-driving-cars-survey}, 2019.

\bibitem{wortham2017}
R.~H. Wortham and A.~Theodorou, ``Robot transparency, trust and utility,''
  \emph{Connection Science}, vol.~29, no.~3, pp. 242--248, 2017.

\bibitem{lee2004}
J.~D. Lee and K.~A. See, ``Trust in automation: Designing for appropriate
  reliance,'' \emph{Human factors}, vol.~46, no.~1, pp. 50--80, 2004.

\bibitem{schwarz2019}
C.~Schwarz, J.~Gaspar, and T.~Brown, ``The effect of reliability on drivers’
  trust and behavior in conditional automation,'' \emph{Cognition, Technology
  \& Work}, vol.~21, no.~1, pp. 41--54, 2019.

\bibitem{hergeth2016}
S.~Hergeth, L.~Lorenz, R.~Vilimek, and J.~F. Krems, ``Keep your scanners
  peeled: Gaze behavior as a measure of automation trust during highly
  automated driving,'' \emph{Human factors}, vol.~58, no.~3, pp. 509--519,
  2016.

\bibitem{korber2018}
M.~K{\"o}rber, L.~Prasch, and K.~Bengler, ``Why do i have to drive now? post
  hoc explanations of takeover requests,'' \emph{Human factors}, vol.~60,
  no.~3, pp. 305--323, 2018.

\bibitem{hergeth2017}
S.~Hergeth, L.~Lorenz, and J.~F. Krems, ``Prior familiarization with takeover
  requests affects drivers’ takeover performance and automation trust,''
  \emph{Human factors}, vol.~59, no.~3, pp. 457--470, 2017.

\bibitem{hoff2015}
K.~A. Hoff and M.~Bashir, ``Trust in automation: Integrating empirical evidence
  on factors that influence trust,'' \emph{Human factors}, vol.~57, no.~3, pp.
  407--434, 2015.

\bibitem{Ayoub2021Modeling}
J.~Ayoub, X.~J. Yang, and F.~Zhou, ``Modeling dispositional and initial learned
  trust in automated vehicles with predictability and explainability,''
  \emph{Transportation Research Part F: Traffic Psychology and Behaviour},
  vol.~77, pp. 102--116, 2021.

\bibitem{merritt2015well}
S.~M. Merritt, D.~Lee, J.~L. Unnerstall, and K.~Huber, ``Are well-calibrated
  users effective users? associations between calibration of trust and
  performance on an automation-aided task,'' \emph{Human Factors}, vol.~57,
  no.~1, pp. 34--47, 2015.

\bibitem{yin2019}
M.~Yin, J.~Wortman~Vaughan, and H.~Wallach, ``Understanding the effect of
  accuracy on trust in machine learning models,'' in \emph{Proceedings of the
  2019 chi conference on human factors in computing systems}, 2019, pp. 1--12.

\bibitem{okamura2020}
K.~Okamura and S.~Yamada, ``Adaptive trust calibration for human-ai
  collaboration,'' \emph{PloS one}, vol.~15, no.~2, p. e0229132, 2020.

\bibitem{ayoub2019}
J.~Ayoub, F.~Zhou, S.~Bao, and X.~J. Yang, ``From manual driving to automated
  driving: A review of 10 years of autoui,'' in \emph{Proceedings of the 11th
  international conference on automotive user interfaces and interactive
  vehicular applications}, 2019, pp. 70--90.

\bibitem{zhang2019}
T.~Zhang, D.~Tao, X.~Qu, X.~Zhang, R.~Lin, and W.~Zhang, ``The roles of initial
  trust and perceived risk in public’s acceptance of automated vehicles,''
  \emph{Transportation research part C: emerging technologies}, vol.~98, pp.
  207--220, 2019.

\bibitem{gold2015}
C.~Gold, M.~K{\"o}rber, C.~Hohenberger, D.~Lechner, and K.~Bengler, ``Trust in
  automation--before and after the experience of take-over scenarios in a
  highly automated vehicle,'' \emph{Procedia Manufacturing}, vol.~3, pp.
  3025--3032, 2015.

\bibitem{beggiato2013}
M.~Beggiato and J.~F. Krems, ``The evolution of mental model, trust and
  acceptance of adaptive cruise control in relation to initial information,''
  \emph{Transportation research part F: traffic psychology and behaviour},
  vol.~18, pp. 47--57, 2013.

\bibitem{guo2020modeling}
Y.~Guo and X.~J. Yang, ``Modeling and predicting trust dynamics in human--robot
  teaming: A bayesian inference approach,'' \emph{International Journal of
  Social Robotics}, pp. 1--11, 2020.

\bibitem{azevedo2020real}
H.~Azevedo-Sa, S.~K. Jayaraman, C.~T. Esterwood, X.~J. Yang, L.~P. Robert, and
  D.~M. Tilbury, ``Real-time estimation of drivers’ trust in automated
  driving systems,'' \emph{International Journal of Social Robotics}, pp.
  1--17, 2020.

\bibitem{luo2020trust}
R.~Luo, J.~Chu, and X.~J. Yang, ``Trust dynamics in human-av (automated
  vehicle) interaction,'' in \emph{Extended Abstracts of the 2020 CHI
  Conference on Human Factors in Computing Systems}, 2020, pp. 1--7.

\bibitem{jian2000}
J.-Y. Jian, A.~M. Bisantz, and C.~G. Drury, ``Foundations for an empirically
  determined scale of trust in automated systems,'' \emph{International journal
  of cognitive ergonomics}, vol.~4, no.~1, pp. 53--71, 2000.

\bibitem{holthausen2020}
B.~E. Holthausen, P.~Wintersberger, B.~N. Walker, and A.~Riener, ``Situational
  trust scale for automated driving (sts-ad): Development and initial
  validation,'' in \emph{12th International Conference on Automotive User
  Interfaces and Interactive Vehicular Applications}, 2020, pp. 40--47.

\bibitem{du2020psychophysiological}
N.~Du, X.~J. Yang, and F.~Zhou, ``Psychophysiological responses to takeover
  requests in conditionally automated driving,'' \emph{Accident Analysis \&
  Prevention}, vol. 148, p. 105804, 2020.

\bibitem{du2020predicting}
N.~Du, F.~Zhou, E.~M. Pulver, D.~M. Tilbury, L.~P. Robert, A.~K. Pradhan, and
  X.~J. Yang, ``Predicting driver takeover performance in conditionally
  automated driving,'' \emph{Accident Analysis \& Prevention}, vol. 148, p.
  105748, 2020.

\bibitem{zhou2020driver}
F.~Zhou, A.~Alsaid, M.~Blommer, R.~Curry, R.~Swaminathan, D.~Kochhar,
  W.~Talamonti, L.~Tijerina, and B.~Lei, ``Driver fatigue transition prediction
  in highly automated driving using physiological features,'' \emph{Expert
  Systems with Applications}, vol. 147, p. 113204, 2020.

\bibitem{Zhou:2019}
F.~Zhou, X.~J. Yang, and X.~Zhang, ``{Takeover Transition in Autonomous
  Vehicles: A YouTube Study},'' \emph{International Journal of Human–Computer
  Interaction}, vol.~0, no.~0, pp. 1--12, 2019.

\bibitem{du2019examining}
N.~Du, J.~Ayoub, F.~Zhou, A.~Pradhan, L.~Robert~Jr, D.~Tilbury, E.~Pulver, and
  X.~J. Yang, ``Examining the impacts of drivers’ emotions on takeover
  readiness and performance in highly automated driving,'' 2019.

\bibitem{zhou2021using}
F.~Zhou, X.~J. Yang, and J.~C. de~Winter, ``Using eye-tracking data to predict
  situation awareness in real time during takeover transitions in conditionally
  automated driving,'' \emph{IEEE Transactions on Intelligent Transportation
  Systems}, 2021.

\bibitem{wang2019exploring}
S.~Wang and Z.~Li, ``Exploring causes and effects of automated vehicle
  disengagement using statistical modeling and classification tree based on
  field test data,'' \emph{Accident Analysis \& Prevention}, vol. 129, pp.
  44--54, 2019.

\bibitem{parasuraman2010complacency}
R.~Parasuraman and D.~H. Manzey, ``Complacency and bias in human use of
  automation: An attentional integration,'' \emph{Human factors}, vol.~52,
  no.~3, pp. 381--410, 2010.

\bibitem{dzindolet2003role}
M.~T. Dzindolet, S.~A. Peterson, R.~A. Pomranky, L.~G. Pierce, and H.~P. Beck,
  ``The role of trust in automation reliance,'' \emph{International journal of
  human-computer studies}, vol.~58, no.~6, pp. 697--718, 2003.

\bibitem{wickens2015engineering}
C.~D. Wickens, J.~G. Hollands, S.~Banbury, and R.~Parasuraman,
  \emph{Engineering psychology and human performance}.\hskip 1em plus 0.5em
  minus 0.4em\relax Psychology Press, 2015.

\bibitem{murtin2018trust}
F.~Murtin, L.~Fleischer, V.~Siegerink, A.~Aassve, Y.~Algan, R.~Boarini,
  S.~Gonz{\'a}lez, Z.~Lonti, G.~Grimalda, R.~H. Vallve \emph{et~al.}, ``Trust
  and its determinants: Evidence from the trustlab experiment,'' 2018.

\end{thebibliography}

\begin{IEEEbiography}[{\includegraphics[width=1in,height=1.55in,clip,keepaspectratio]{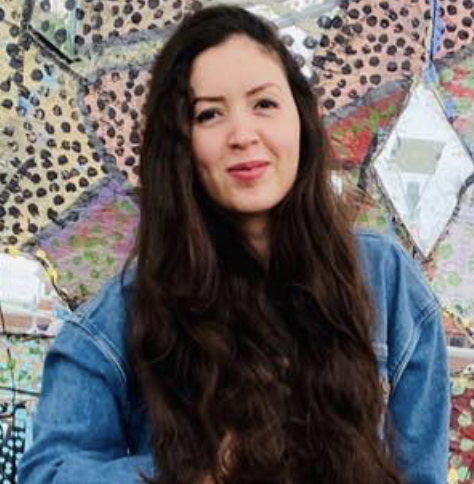}}]{Jackie Ayoub received her B.E. degree in mechanical engineering from Notre Dame University, Lebanon, in 2016 and her master degree in Industrial and Systems Engineering from University of Michigan, Dearborn, in 2017. She is currently pursuing her Ph.D. in Industrial and Systems Engineering in the University of Michigan, Dearborn. Her main research interests include human-computer interaction, human factors and ergonomics, and sentiment analysis.}
\vspace{-20pt}
\end{IEEEbiography}

\begin{IEEEbiography}[{\includegraphics[width=1in,height=1.55in,clip,keepaspectratio]{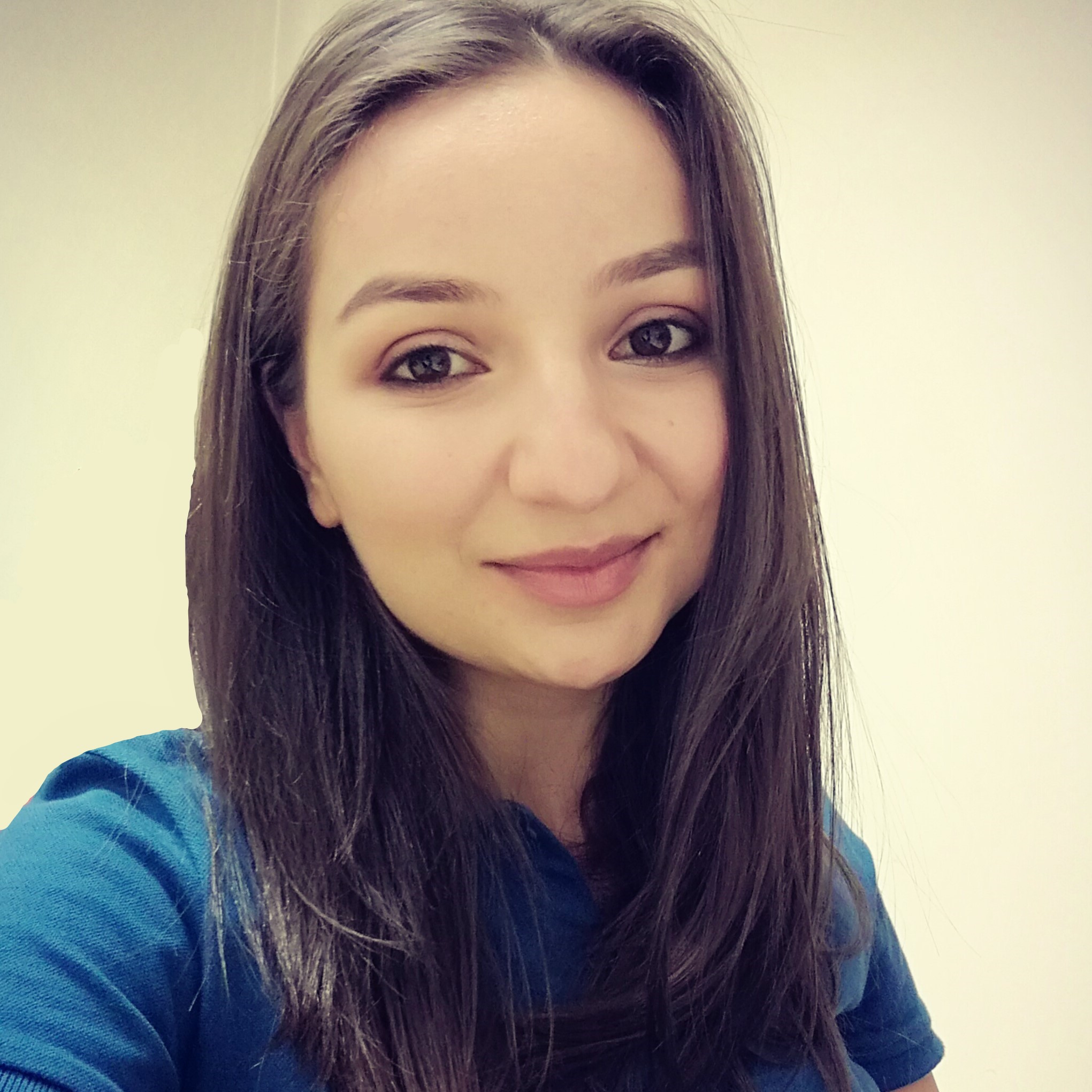}}]{Lilit Avetisyan received her B.E. degree in 2017 and MS degree in 2019 in Information Security from the National Polytechnic University of Armenia. She is currently pursuing her Ph.D. degree in Industrial and Systems Engineering at the University of Michigan, Dearborn. Her main research interests include human-computer interaction, explainable artificial intelligence and human-centered design.}
\vspace{-20pt}
\end{IEEEbiography}

\begin{IEEEbiography}[{\includegraphics[width=1in,height=1.55in,clip,keepaspectratio]{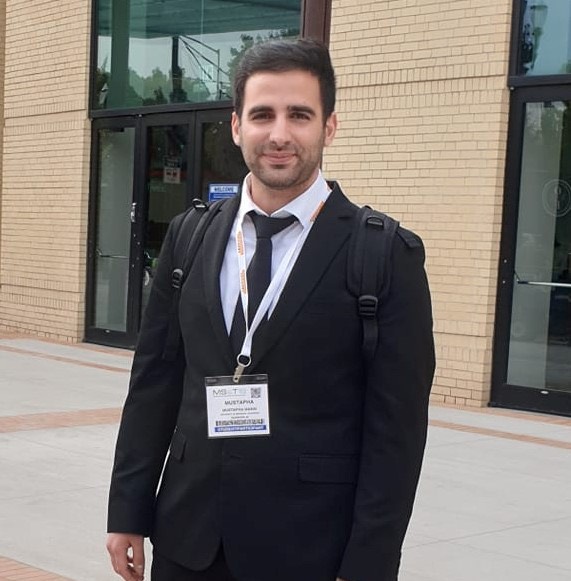}}]{Mustapha Makki received his bachelor’s degree in mechanical engineering from Texas A\&M University in 2014, and his master's degree from the American University of Beirut in 2017. He is currently pursuing his Ph.D. in industrial and Systems Engineering at the University of Michigan, Dearborn. His main research interests include experimental characterization and modeling of the mechanical and damage behavior of ductile metals and polymers and introducing virtual and mixed reality tools to promote active learning in engineering.}
\vspace{-20pt}
\end{IEEEbiography}

\begin{IEEEbiography}[{\includegraphics[width=1in,height=1.55in,clip,keepaspectratio]{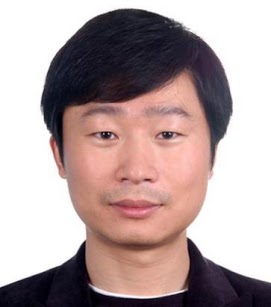}}]{Dr. Feng Zhou received the Ph.D. degree in Human Factors Engineering from Nanyang Technological University, Singapore, in 2011 and Ph.D. degree in Mechanical Engineering from Gatech Tech in 2014. He was a Research Scientist at MediaScience, Austin TX, from 2015 to 2017. He is currently an Assistant Professor with the Department of Industrial and Manufacturing Systems Engineering, University of Michigan, Dearborn. His main research interests include human factors, human-computer interaction, engineering design, and human-centered design.}

\end{IEEEbiography}

\end{document}